\documentclass[a4paper,11pt,twocolumn]{article}

\usepackage{graphicx} 
\usepackage{dcolumn} 
\usepackage{bm} 
\usepackage{gensymb} 
\usepackage{amsmath} 
\usepackage{float} 
\usepackage{authblk} 
\usepackage{abstract} 
\usepackage{tabularray} 
\usepackage[margin=1.5cm]{geometry} 
\def\bx{{\bf x}}              %
\def\bu{{\bf u}}              %
\newcommand{\eq}[1]{Eq.~(\ref{#1})}
\newcommand{\be}{\begin{equation}} 
\newcommand{\ee}{\end{equation}} 
\newcommand{\bea}{\begin{eqnarray}} 
\newcommand{\eea}{\end{eqnarray}}

\title{Droplet spreading in a wedge: A route to fluid rheology for power-law liquids}

\author[1]{Marcel Moura}
\author[2]{Vanessa Kern}
\author[1,3]{Knut J{\o}rgen M{\aa}l{\o}y}
\author[2,4]{Andreas Carlson}
\author[1,5]{Eirik G. Flekk{\o}y\thanks{Corresponding author: flekkoy@fys.uio.no}}

\affil[1]{\small PoreLab, the Njord Center, Department of Physics, University of Oslo, NO--0316 Oslo, Norway}
\affil[2]{\small Department of Mathematics, University of Oslo, NO--0316 Oslo, Norway}
\affil[3]{\small PoreLab, Department of Geoscience and Petroleum, Norwegian University of Science and Technology, NO--7031 Trondheim, Norway}
\affil[4]{\small Department of Medical Biochemistry and Biophysics, Umeå University, Sweden}
\affil[5]{\small Department of Chemistry, Norwegian University of Science and Technology, NO--7491 Trondheim, Norway}

\date{\today}

\begin{document}
\twocolumn[
    \begin{@twocolumnfalse}
        \maketitle
        \vspace{-0.8cm}
        \begin{abstract}
                Measuring the rheology of liquids typically requires precise control over shear rates and stresses. Here, we describe an alternative route for predicting the characteristic features of a power-law fluid by simply observing the capillary spreading dynamics of viscous droplets in a wedge-shaped geometry. In this confined setting, capillary and viscous forces interact to produce a spreading dynamics described by anomalous diffusion, a process where the front position grows as a power-law in time with an exponent that differs from the value $1/2$ found in classical diffusion. We derive a nonlinear diffusion equation that captures this behavior, and we show that the diffusion exponent is directly related to the rheological exponent of the fluid. We verify this relationship by using both experiments and simulations for different power-law fluids. As the predictions are independent from flow-specific details, this approach provides a robust tool for inferring rheological properties from the spreading dynamics.
        \end{abstract}
        \vspace{1cm} 
    \end{@twocolumnfalse}
]

The spreading of droplets on solid surfaces is a ubiquitous phenomenon, observable in everyday events such as raindrops falling on a window, as well as in a variety of biological, geological, and physical systems. In these contexts, it can serve as a mechanism for delivering nutrients to living cells \cite{steudle1998}, oxidating substances to minerals \cite{jamtveit99}, or forming fluid pathways within complex geometries \cite{tuller1999,meakin2009}. The spontanoeus spreading of fluids depends on the wetting properties and geometry of the medium through which it spreads \cite{concus1969behavior,PhysRevE.85.045302,RevModPhys.57.827,D3SM00715D,annurev:/content/journals/10.1146/annurev-fluid-011212-140734}. However, it also depends crucially on the rheology of the liquid, as is well known from the everyday application of non-Newtonian fluids, such as toothpaste, corn starch, paint, yogurt and shampoo. Controlling fluid rheology is also topical in industrial applications and in food science \cite{irgens2016,bingham1917,coussot1999,schowalter1978,heldman2019,benezech1994,bertsch2019,phillips2009}, where it covers everything from cooking recipes to consumer satisfaction \cite{deblais2021}. 
 In this study, we show that the spreading dynamics of a droplet of a power-law liquid can be used to predict the fluid's rheological properties.

The rheology of a power-law fluid is characterized by the fact that the shear stress is given by a power of the strain rate. The spreading of such a liquid in a wedge-shaped geometry is anomalous in the sense that it is described in the same way 
as anomalous diffusion. Mathematically, it is captured by a non-linear diffusion equation that yields analytical solutions with a spreading rate
which is given by an anomalous diffusion exponent \cite{Bouchaud,flekkoy21,hansen2011,article}.
As an interesting by-product, we obtain a relationship between the exponent governing the rheology and the diffusion exponent.  As this relationship depends only on the conservation of fluid mass and a non-linear Darcy law, which relates the volume flux to the pressure gradient, it is independent of other details of the flow structure. The rheological exponent may thus be obtained solely on the basis of the spontaneous (capillary driven) fluid motion, without any type of force- or shear-rate control, as needed in commercial rheometers.

\begin{figure}
\begin{center}
\includegraphics[width=\columnwidth]{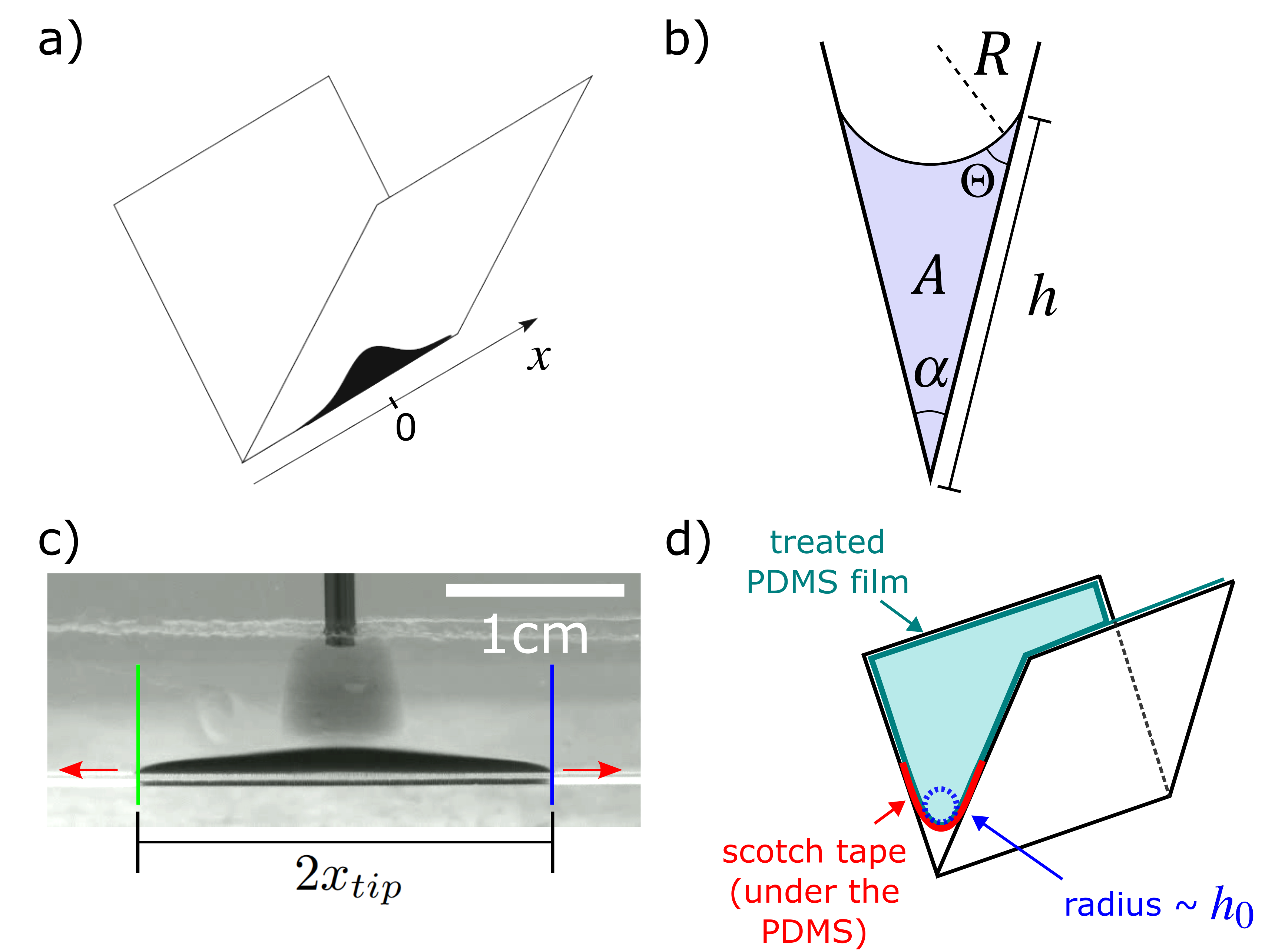}
\caption[]{a) Liquid droplet spreading in a wedge. b) The cross-sectional area $A$ and the local height $h$ of the spreading droplet, which both vary along the wedge. Here, $\alpha= 30^\circ$ is the opening angle of the wedge, and $\Theta$ is the contact angle at the liquid-solid-air contact line. c) Snapshot of the experiment, where the green and blue vertical lines denote the left and right tip position, and the red arrows indicate the spreading direction. d) Diagram illustrating the Scotch tape (red) and PDMS layer (cyan) employed, which lead to an effective finite radius of curvature $h_0$ in the corner (exaggerated in the figure for clarity). Note that the main difference between the idealized (sharp) wedge and the experimental setup is a finite curvature where the plates meet.}
\label{fig:experimentalsetupcombined}
\end{center}
\end{figure}

The analysis and findings presented here complement the extensive literature on anomalous diffusion. Pattle \cite{pattle59} noted, already in 1959, that the seepage of a liquid into a cloth or porous body could be described by a nonlinear diffusion equation, yielding sub-diffusive spreading of the fluid. In the 1980s, there was a large effort to understand the nature of anomalous diffusion, particularly in the context of transport in disordered and porous media. In porous media flows, strong confinement often gives rise to regimes where transport occurs through corners and thin films. In these cases, the balance between capillary and viscous forces plays a key role and can enable persistent, long-range transport \cite{tuller2001,hoogland2016,moura2019,reis2023,concus1969behavior} such as when connecting otherwise disconnected fluid clusters \cite{savins1969,cannella1988,sochi2008,eberhard2020,lima2022,lanza2022,an2022,fyhn2023,zhang2023}. Our solution, which pertains to a viscous flow driven by capillarity as a droplet spreads in an idealized (sharp) wedge, see Fig.~\ref{fig:experimentalsetupcombined}a), with a constant solid-liquid contact angle, may thus serve as a baseline model for more complex flow geometries. 

A key finding in our study is the fact that the position of the tip of a droplet spreading in a wedge geometry scales with time as $ x_{tip}\propto t^{\tau}$, where the exponent $\tau$ is directly linked to the rheological exponent $n$ characterizing the power-law fluid. By contrast, when the spreading of a shear-thinning fluid happens in  radial fashion on a plane \cite{starov2003,rafai05},  
rather than along a corner line, the exponent governing the spreading rate has been shown to depend very weakly on the rheological exponent \cite{rafai05}. 
Our theoretical prediction and measurements of the Newtonian spreading exponent $\tau \approx 0.4$ are consistent with the value $\tau  \approx 1/3$ that has been measured for Newtonian flow of different liquids that imbibe along a corner  from an infinite reservoir \cite{ponomarenko2011,wijnhorst20}. This suggests that the hydrodynamics is strongly governed by the dynamics at the tip and less sensitive to the boundary condition where the fluid is supplied.

A droplet of a power-law fluid spreading in a wedge
is characterized by the stress-strain relationship $\sigma \propto \dot{\gamma}^n$ over a certain  range of the strain rate $\dot{\gamma}$, which is frequently described by the Ostwald–de Waele model \cite{schowalter1978}. The viscosity is then given by 
\be
\mu  =\eta_0 \dot{\gamma}^{n-1} = \mu_0 \left( \frac{\dot{\gamma}}{\dot{\gamma}_0} \right)^{n-1}\:,
\ee
where $\mu_0$ and $\mu$ are in units of $\mathrm{Pa\,s}$, the exponent $n=1$ corresponds to a Newtonian fluid, and $\dot{\gamma}_0$ is a reference strain rate at which the liquid is non-Newtonian. We shall take 
$\dot{\gamma}_0 = 10\,\mathrm{s}^{-1}$
for the fluids employed in this study (see the rheological characterization of the fluids in the Supplemental Material).
Many polymer melts and solutions exhibit a value of $n$ in the range  
$0.3-0.7$, depending upon the concentration and molecular weight of the  
polymer \cite{deshpande10,carreau1972}.  
Even smaller values of the power-law exponent ($n=0.1 - 0.15$)  
are encountered with fine particle suspensions like kaolin-in-water,  
bentonite-in-water, etc.  
By using polyacrylamide solutions of different concentrations, Ansari et al.
\cite{ansari20} measured a range of power-law indices $n = 0.26-0.47$, which were obtained for shear rates in the range $10-1000\,\mathrm{s}^{-1}$. See also Jouenne and Levache\cite{jouenne2020} for a comprehensive dataset on the rheology of acrylamide-based polymer solutions.

 We shall obtain a prediction for the height $h(x,t)$ of a droplet of a power-law fluid when spreading along the wedge geometry, as shown in Fig.~\ref{fig:experimentalsetupcombined}.  Our power-law liquid  wets the surface inside the  wedge and  spreads due to the capillary forces, as illustrated in  Fig.~\ref{fig:experimentalsetupcombined}a). Its shape is described by the height $h(x,t)$. When the wetting angle $\Theta < (\pi -\alpha)/2$, $\alpha$ being the opening angle of the wedge (see Fig.~\ref{fig:experimentalsetupcombined}b)), the capillary pressure will be negative, and the minimal energy will be reached when the solid-liquid contact area is maximized. This occurs when the liquid spreads out as far as possible along the wedge. The negative capillary pressure drives the flow and increases in magnitude towards the tips of the drop as $1/h$, since the radius of curvature of the liquid-air interface $R \propto h$, see Fig.~\ref{fig:experimentalsetupcombined}b). In Fig.~\ref{fig:experimentalsetupcombined}c) we show a snapshot of the experiment, where the red arrows denote the spreading direction. It is worth noting that in the experimental setup, the wedge is not perfectly sharp as in the idealized model. Instead, a small radius of curvature $h_0$ exists along the bottom edge, see Fig.~\ref{fig:experimentalsetupcombined}d). This curvature arises from the procedure employed to build the setup, which involves coating the glass slides that form the wedge walls with a PDMS film. Further details are provided in the Supplemental Material.

In the Supplemental Material we show that the volumetric flow rate $q$ across a surface normal to the corner line may be written 
\be
q  =  Q_n\left( \frac{-\partial P/\partial x }{{\eta_0} }\right)^{1/n} 
h^{3+1/n}\:,
\label{kuyf2}
\ee
where $P$ is the capillary pressure, $x$ is the coordinate along the wedge, $h(x)$ is the height of the liquid in the
normal direction, and
\be
Q_n = \frac{2^{\frac{1+n}{2n}} n^2}{(2n+1)(3n+1)} \left( \frac{\alpha}{2} \right)^{\frac{2n+1}{n}}
\ee
is a dimensionless average of the flow field, which is  dependent on our choice of flow geometry, but not its length scale. See also Guyon et al.\cite{guyon2014} for further details on lubrication flows and complex fluids.

In order to obtain an analytical solution, our theoretical approach ignores: 1) the effects of gravity and inertia, 2) the finite initial droplet width and the interface curvature in the $x$-direction, 3) the cut-offs in shear rates where the non-Newtonian behavior becomes Newtonian, 4) the finite curvature of the corner geometry, and 5) dynamic variation of the contact angle $\Theta$ and pinning effects.

\begin{table}[t]
\setlength{\tabcolsep}{0.5pt} 
  \centering
  \begin{tabular}{|c|c|c|c|c|}
      \hline
       XG [$\mathrm{g/L}$] & $u_m$ [$\mathrm{m/s}$] & $\nu^*$ [$\mathrm{m^2/s}$] & $\dot{\gamma}_{\text{tip}}$ $(\dot{\gamma}_c)$ [$\mathrm{s^{-1}}$] & Re \\   
        \hline
0     & $1\,\times\,10^{-2}$    & $5\,\times\,10^{-6}$   & 70 (0)       & $2\,\times\,10^{-2}$ \\
0.5   & $1\,\times\,10^{-2}$    & $1\,\times\,10^{-5}$   & 70 (3)     & $1\,\times\,10^{-2}$ \\
1     & $6\,\times\,10^{-3}$    & $4\,\times\,10^{-5}$   & 40 (2)     & $1\,\times\,10^{-3}$ \\
2     & $1\,\times\,10^{-3}$    & $2\,\times\,10^{-4}$   & 7 (0.8)      & $5\,\times\,10^{-5}$ \\
3     & $5\,\times\,10^{-5}$    & $2\,\times\,10^{-3}$   & 0.35 (0.2)   & $1\,\times\,10^{-7}$ \\
4     & $4\,\times\,10^{-5}$    & $4\,\times\,10^{-3}$   & 0.3 (0.1)    & $1\,\times\,10^{-7}$ \\
6     & $2\times10^{-5}$    & $2\,\times\,10^{-2}$   & 0.15 (0.08)  & $1\,\times\,10^{-8}$ \\

       \hline
  \end{tabular}
  \caption{Characteristic values of the flow velocity $u_m$, shear rate at the droplet tips $\dot{\gamma}_{tip}$, and Reynolds number $\mathrm{Re}$. 
  The threshold shear rate $\dot{\gamma}_c$, at which non-Newtonian behavior sets in, is shown in parentheses alongside $\dot{\gamma}_{tip}$. 
  The values of $\dot{\gamma}_c$ are obtained from Fig.~\ref{fig_rheology}, which also provides the corresponding kinematic viscosity values $\nu^* = \eta / \rho$. 
  The Reynolds number is calculated as $\mathrm{Re} = u_m L_{\mathrm{eff}}/\nu^*$, where the effective length $L_{\mathrm{eff}} \approx 10^{-5}\,\mathrm{m}$.}
  \label{table}
\end{table}

 Approximating the contact angle by $\Theta = 20^\circ$, the curvature of
the liquid interface $1/R$ sets the  capillary pressure difference across this interface to be
\be
 P(x) =  -\frac{\cos (\Theta + \alpha /2)\sigma}{h(x)\sin (\alpha /2)}\:,
 \label{iglih}
\ee
where $\sigma$ is the liquid-air surface tension and we have applied a simple geometric argument
to replace $R$ by $h$ and the relevant angles shown in Fig.~\ref{fig:experimentalsetupcombined}b).

The relative magnitude of gravitational to capillary forces is estimated as the ratio  between the hydrostatic pressure drop
$\rho g h$ and the capillary pressure drop given in  \eq{iglih}. Ignoring the cosine term in this expression yields the ratio
\be
\frac{\rho g h^2}{\sigma}
\sin (\alpha/2)\approx 0.15
\ee
for all the experiments. The fraction is well below unity, justifying the neglect of gravity in the theory. Here we have set $h = 2\,\mathrm{mm}$,  which is a typical maximum value during the observation time of the experiments, and the opening angle of the wedge is $\alpha= 30^\circ$. The fluid density $\rho$ and surface tension $\sigma$ do not differ significantly from pure water. 

In order to estimate the relative importance of inertial forces, we calculate the Reynolds number for each experiment, that is, the ratio of the steady inertial and viscous terms of the Navier-Stokes
equation. The quantity $u_m$ represents a characteristic mean tip velocity, measured as the ratio $x_{tip}/t$ in the late-time regime of the experiment, where the scaling $x_{tip}\propto t^{\tau}$ holds. While the actual velocity decreases over time, $u_m$ serves as a useful reference for the typical velocity observed during the experiment. Calculating a relevant Reynolds number requires some
care, as the viscous term $\propto \nabla^2 u$ is dominated by the shear forces from the wall and thus governed
by the width of the droplet, while the inertial term $\propto\bu \cdot \nabla \bu $ is governed by variations along the flow and hence the droplet half-width $x_{tip}$. As a result we define the relevant Reynolds number as
\be
\mbox{Re}= \frac{u_mL_{eff}}{\nu^*}
\ee
with the effective length  $L_{eff}= (\alpha h/2)^2/x_{tip}\approx 10^{-5}\,\mathrm{m}$ when the typical value for the droplet half-width is $x_{tip} \approx 2\,\mathrm{cm}$. Here, $\nu^*$ is the kinematic viscosity which we shall
extract from figure \ref{fig_rheology} as the value in the Newtonian regime. 
In Table \ref{table}
the values of Re for the different experiments are given, showing that inertia is indeed negligible, and more so for the higher XG-concentrations.

Assuming the fluids to be in the non-Newtonian regime  is clearly not correct everywhere.
Since the shear rate will be close to zero at $x=0$, there will always be regions of slow Newtonian flow.
However, since the capillary forces and viscous dissipation take on their maximum values at $x=\pm x_{tip}$, the spreading process is likely to be strongly dominated by the
hydrodynamics right at the droplet tips. 
For that reason, we compare the shear rate there to the critical crossover shear rate $\dot{\gamma}_c$, where the rheology becomes Newtonian. 
Taking $h_0$ to be the only relevant length scale for the flow at the tips, we estimate the  tip shear rate as $\dot{\gamma}_{tip}=u_m/(\alpha h_0/2)$. 
Table \ref{table}   shows the typical shear rate, $\dot{\gamma}_{tip}$,
and how it compares with the crossover shear rates $\dot{\gamma}_c$ obtained from
figure \ref{fig_rheology}.
Note that in all cases $\dot{\gamma}_{tip} > \dot{\gamma}_c$, although
only by a small margin for the largest concentrations. We will include the influence of a finite corner curvature in our numerical solutions by setting the volumetric flow rate to $q = 0$ when $h$ falls below a threshold $h_0$, see Fig.~\ref{fig:experimentalsetupcombined}d).

It is the variations in $h(x)$ that will cause a pressure gradient
along the wedge, and we may invoke \eq{kuyf2} to get the mean flow.
 The cross-sectional  area in the direction normal to $x$, $A(x) \approx (\alpha /2 )h^2$, and the amount of liquid volume is a conserved quantity, so that
\be
\frac{\partial A }{\partial t} + \frac{\partial q}{\partial x}  =0 \:.  
\label{conserved}
\ee  
Expressing $h$ by $A$ in \eq{kuyf2} and using \eq{iglih}, the volume conservation may then be written on
the form
\be
\frac{\partial A }{\partial t} = - D_0 \frac{\partial}{\partial x}  \left( 
A^{3/2-1/n}  \left( -\frac{\partial A  }{\partial x}    \right)^{1/n}   \right)\:,
\label{conserved2}
\ee 
where the `diffusion coefficient'
\be
D_0  = \left( \frac{2}{\alpha}\right)^{3/2} \left(\frac{\sigma}{\eta_0} \frac{\cos (\Theta + \alpha /2 )}{2\sin (\alpha /2)} \right)^{1/n}Q_n
\label{eq:D0}
\ee
has dimensions of $\mathrm{m}^{1/n}/\mathrm{s}$  .

Since the solution $A(x,t)$ is symmetric around $x=0$, we will only consider $x>0$,
so that ${\partial P}/{\partial x}<0$.
By taking the initial condition to be $A(x,0)=V_0 \delta (x)$, where $V_0$ is the
droplet volume, we shall follow \cite{flekkoy21} and seek a scaling solution of the form
\be
A(x,t) = \frac{p(u)}{f(t)} \:,
\label{kuykugg}
\ee
where
\be
u=\frac{x}{f(t)} \:.
\label{juytf}
\ee
This solution has the property that 
$
\int dx A(x,t) = \int du \; p(u)=V_0\:. 
\label{7fkuy}
$
In order to insert \eq{kuykugg} in \eq{conserved2}, we need the derivative,
${\partial A}/{\partial t}  = -({f'(t)}/{f(t)^2})  d(up)/du$, which allows us to write \eq{conserved2}  in the form 
 
\be
\frac{f'(t)}{D_0f^{-1/2-1/n}}= \frac{(p^{3/2-1/n} (-dp/du)^{1/n})'}{d(u p)/du}
= \lambda \:,
\label{jhud}
\ee
where we have separated the $u$ and $t$-dependent terms with
the separation constant $\lambda$. 
Integration of the $f(t)$-part  of \eq{jhud} is straightforward and yields
\be
f(t) =  \left( \frac{\lambda D_0t}{\tau}  \right)^{\tau} \:,
\label{kuyf4}
\ee
with
\be
\tau =  \frac{2n}{2+3n}\:.
\label{tau}
\ee

The $p$-part of \eq{jhud}  takes the form,
$  (d/du)( p^{3/2-1/n} (-dp/du)^{1/n}) - \lambda u p ) =0 $,
which may be integrated to give
$ p^{3/2-1/n} (-p'(u))^{1/n} - \lambda u p  = K $.
Since, by symmetry, $p'(0)=0$ and $p(0)$ must be finite, the
integration constant  $K=0$, and the
above equation may be rearranged to give $
p^{n/2-1}p' = -\lambda^n u^n $.
This may be integrated to give
\be
p(u) = \lambda^2 \left( \frac{n}{2(n+1)} \right)^{2/n} 
( L^{n+1} - u^{n+1} )^{2/n}\:, 
\label{uygo}
\ee
where the normalization
condition yields the integration constant $L=\lambda^{-\tau} (V_0/A_n)^{\tau /2}$
with $A_n= 2(n/(2(n+1)))^{2/n}\int_0^1 dy (1-y^{n+1})^{2/n}$.
Inserting this $L$-value in $A(x,t) = p(u)/f(t)$, $\lambda$ cancels out, and so we are
free to choose $\lambda = \dot{\gamma}_0 /D_0$ in order to make 
\be f(t)= \left( \frac{\dot{\gamma}_0 t}{\tau} \right)^{\tau}
\ee
dimensionless. Then, $L(D_0)=((D_0 / \dot{\gamma}_0)  (V_0/A_n)^{1/2})^{\tau }$, where the dimension of $u$ is $\mathrm{m}$ and that of $p(u)$ is $\mathrm{m}^2$.
Inserting $\lambda$ in \eq{uygo} finally yields
\be
p(u) = \left( \frac{\dot{\gamma}_0}{D_0} \right)^2 \left( \frac{n}{2(n+1)}\right)^{2/n}  ( L^{n+1}(D_0) - u^{n+1})^{2/n}\:.
\label{kuyy}
\ee
Requiring that $p(u)$ be real restricts this solution to the 
$x$-domain where  $ L^{n+1} - u^{n+1}>0$, or
$|u|<L$. Outside 
\be
x_{tip}(t)  = L  {f(t)}  =   \left( \left( \frac{V_0}{A_n} \right)^{1/2}\frac{D_0t}{\tau} \right)^{\tau} \:,
\label{kuyftd}
\ee
$A(x,t)$ then vanishes exactly, so that 
$2x_{tip}(t) $ is the extent of the droplet. 

The fact that $ x_{tip}\propto t^{\tau}$
is a key result as it provides the link between the spreading rate and the diffusion exponent $\tau$, and by \eq{tau}, the rheological exponent $n$. The $\tau = 0.4$ value for $n=1$ agrees with the result found by Hansen et al. \cite{hansen2011}, who solved \eq{conserved2} in the special case of a Newtonian liquid.
The $\tau$-exponent  is simply obtained by plotting the experimental $x_{tip}(t)$-values on a log-log plot. 
Having obtained the $\tau$ and $n$-values from $x_{tip}(t)$ in this way, the 
scaling function $f(t)$ is known and may be used to obtain 
a data collapse for the experimental values of $A(x,t)f(t)$
versus $x/f(t)$ as predicted by theory. 
With this experimental data collapse, fitting \eq{kuyy} to the data using
$D_0$ as a fitting parameter gives $D_0$.
The  rheological prefactor $\mu_0$ may then in principle  be obtained by 
 solving 
\eq{eq:D0} with respect to $\mu_0 = \eta_0 \dot{\gamma}_0^{n-1}$ as
\be
\mu_0  = \left( 
\frac{2}{\alpha}\right)^{3n/2}  
\frac{\sigma \cos (\Theta + \alpha /2 )}{2\sin (\alpha /2)} 
\left( \frac{Q_n}{D_0} \right)^n \dot{\gamma}_0^{n-1},
\label{eq:eta0} 
\ee
provided a constant contact angle.  However, as is well known, the static contact angle, which we measure 
to be $\Theta \approx 20^\circ$, will in general be replaced by a dynamic contact angle $\Theta_{dyn}$
once the contact line is moving. Hoffmann \cite{hoffmann75} showed that the difference between $\Theta_{dyn}$ and $\Theta$ is governed
by the local capillary number Ca via a general scaling relation. 
Later, Sheng  \cite{sheng92} and Kokko-Latva and Rothmann \cite{kokko07} showed that this relation 
is well approximated by $\cos (\Theta) - \cos (\Theta_{dyn} ) \propto \mbox{Ca}$ with a prefactor 
of order unity.   
The capillary number Ca is the ratio of viscous to capillary forces, so in our case Ca$\sim$1, since these are the dominating  forces that govern the flow, in particular close to $x_{tip}$. 

Also, since the analytic theory ignores the corner curvature $h_0$, an experimentally measured value of $h(x,t)$ (the distance from the curved corner to the meniscus) will be associated
with an overestimated value of the capillary pressure that drives
the flow. This will predict a faster overall spreading rate than what is actually observed. This effect thus has the opposite influence
of the dynamic contact angle increasing above the static one, which will lead to a slower spreading. For these reasons $\mu_0$ cannot be obtained from \eq{eq:eta0}
unless $\Theta_{dyn}$ is measured independently and used instead of $\Theta$, and the corner is made sufficiently sharp.

\begin{figure}[t]
\begin{center}
\includegraphics[width=0.9\columnwidth,angle=0]{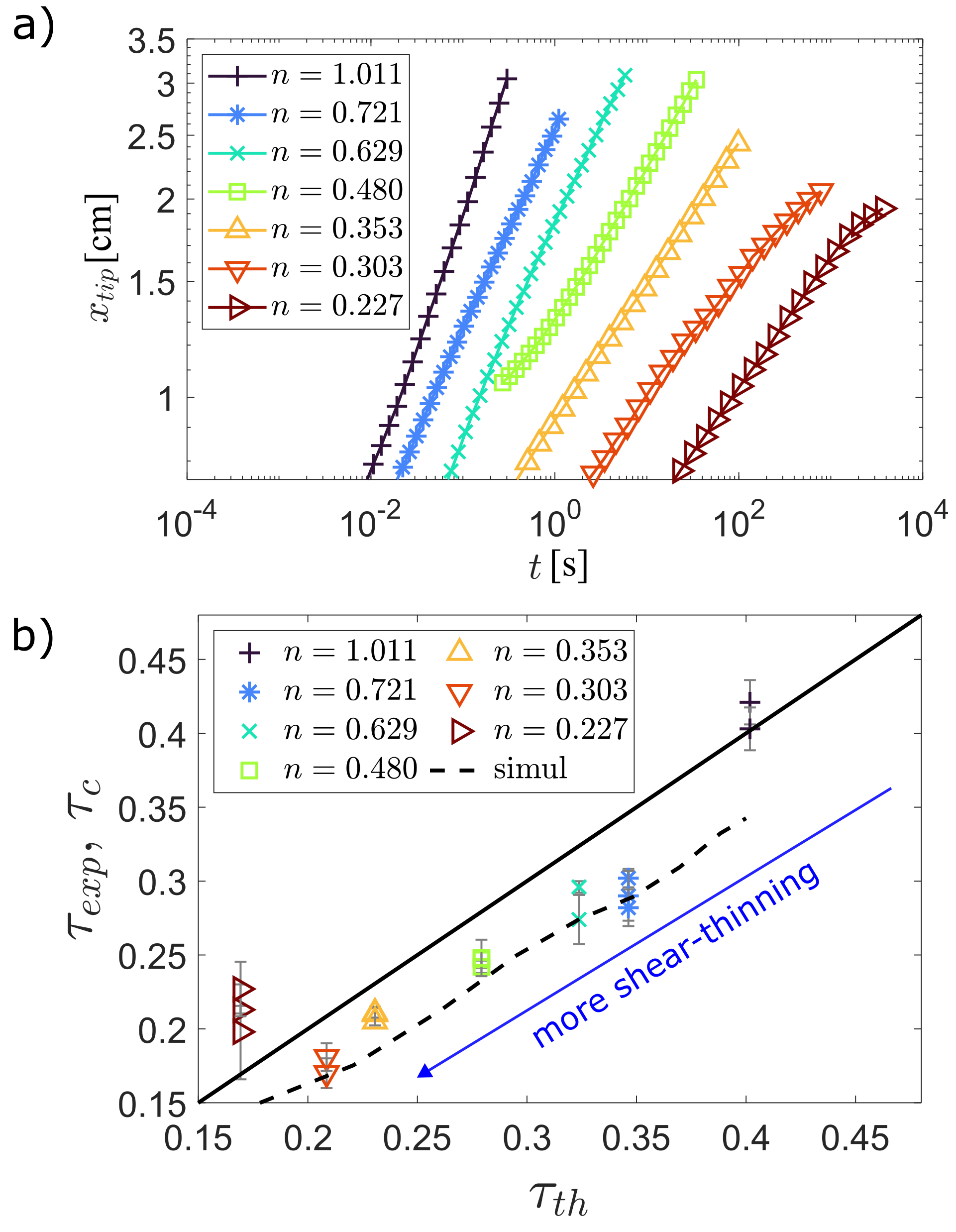}
\caption[]{a) Experimentally measured droplet spread  $x_{tip}(t)$. The curves are shifted horizontally by arbitrary values to aid visualization. The leftmost curve corresponds to the Newtonian fluid, while the  others are from XG fluids with XG concentrations of 0.5, 1, 2, 3, 4 and 6 g/L increasing from left to right. b) Comparison between the experimental exponent $\tau_{exp}$ (symbols) and numerical exponent $\tau_{c}$ (dashed line), as well as the theoretical slopes $\tau_{th}$ computed from \eq{tau} (full line). The numerical results take the finite corner curvature $h_0 = 0.6\,\mathrm{mm}$
 and initial profile half-width $w_0 = 0.8\,\mathrm{mm}$ into account. The errorbars account for deviations from a perfect power-law fit from the data in a).} 
\label{fig:xtip}
\end{center}
\end{figure}

The mathematical model can be solved numerically, and the solutions provide testable predictions. To test these predictions, we design an experimental system consisting of two thin transparent plates, arranged at an angle $\alpha= 30^\circ$ and securely held by a support mechanism (see Fig.~\ref{fig:experimentalsetupcombined}). Standard microscope glass slides (dimensions $76\,\mathrm{mm} \times 26\,\mathrm{mm} \times 1\,\mathrm{mm}$) were used. One critical experimental aspect was to ensure a high level of wettability with the wedge and minimize impurities that may affect the contact-line motion. As such, the plates were coated with a thin polymeric layer of PDMS, modified with a hydrophilic agent (methyl-terminated poly(dimethylsiloxane-b-ethylene oxide)). The coating was further treated with nitrogen plasma to enhance the wettability \cite{tan2010}. We tested a range of different fluids, including mixtures of xanthan gum (XG) \cite{whitcomb1977} and water at varying concentrations, as well as a glycerol-water mixture to represent the Newtonian case. The XG concentration in water was systematically varied from $0.5 \,\mathrm{g/L}$ to $6 \,\mathrm{g/L}$. For further details about the coating procedure and fluids preparation, see the Supplemental Material.

\begin{figure}[t]
\begin{center}
\includegraphics[width=1\columnwidth,angle=0]{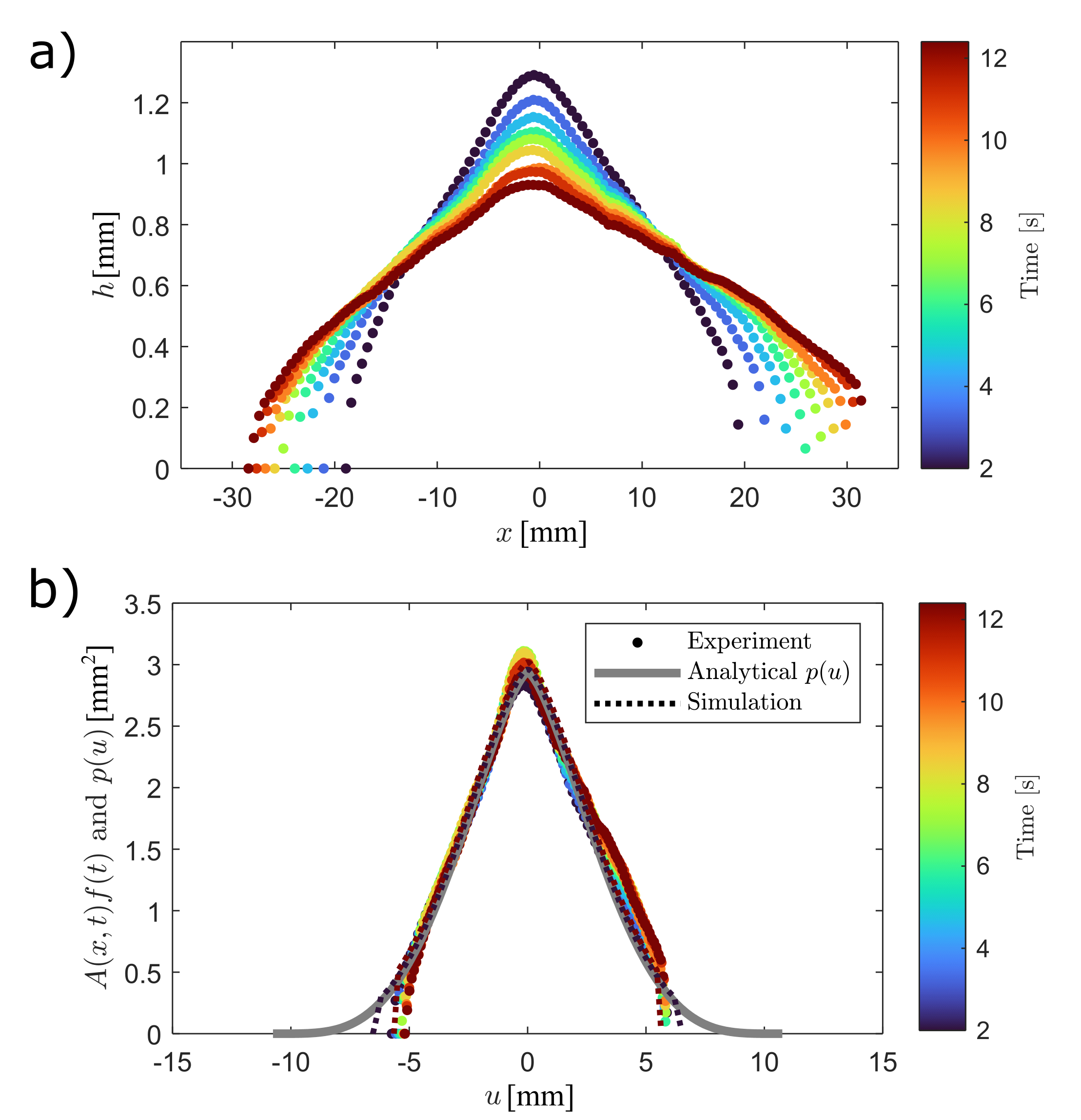}
\caption[]{a) Droplet profiles for an experiment with XG concentration of $2 \,\mathrm{g/L}$ for 9 different times in the range $t = 2.0\text{–}12.4\,\mathrm{s}$ shown in the colorbar. b) The  dots show the experimental values of  $A(x,t)f(t)$. The analytical result for the function $p(u)$ of \eq{kuyy} using the value $n= 0.48$ is plotted as a gray line, while the simulations that take the finite corner curvature $h_0 = 0.6\,\mathrm{mm}$
and intitial profile half-width $w_0 = 0.8\,\mathrm{mm}$,  are shown  for the initial and final  times. The diffusion constant resulting from the data fit is $D_0 = 200\,\mathrm{mm}^{1/n}/\mathrm{s}$.}
\label{fig:datacollapse}
\end{center}
\end{figure}

In the experiment the wedge's corner is not infinitely sharp, rather it has a radius of curvature $h_0 \approx 0.6\,\mathrm{mm}$ due to the finite thickness of the PDMS coating, see Fig.~\ref{fig:experimentalsetupcombined}d). To mimic the experiments, we solve
\eq{conserved} numerically, introducing a Gaussian initial profile for $h(x,t)$
and a cut-off corresponding to the corner curvature.
This cut-off arises because the capillary pressure changes sign
approximately at the
point  where $h(x,t)$ becomes smaller than $h_0$, see Fig.~
\ref{fig:experimentalsetupcombined}d). We set the radius of curvature $h_0 = 0.6\,\mathrm{mm}$ and the intitial profile half-width $w_0 = 0.8\,\mathrm{mm}$ throughout in the numerical
calculations, similar to the experiments. 
This may be represented mathematically 
 by imposing $q=0$ in \eq{conserved}, where $h(x,t) <h_0$. Equation (\ref{conserved2}) can then be integrated numerically using centered
 spatial derivatives and an explicit first-order time stepping scheme.

 Fig.~\ref{fig:xtip}a) shows the temporal evolution of the droplet spread $x_{tip}$, measured as half of the distance between the left and right tips of the droplet, marked in Fig.~\ref{fig:experimentalsetupcombined}b), for seven experiments going from the Newtonian case (leftmost curve) to the most shear-thinning case (rightmost curve). Arbitrary horizontal shifts have been applied to the curves for better visualization. The linear  range in the $x_{tip}$ plot allows the 
determination of $\tau$ by use of \eq{kuyftd}. Fig.~\ref{fig:xtip}b) shows a comparison between the measured $\tau$-values against their theoretical predictions. The experimentally and numerically measured exponents are denoted $\tau_{exp}$ and $\tau_{c}$, and the theoretical values $\tau_{th}$ are computed  from \eq{tau}. The errorbars shown are due to the uncertainty in the fitting of the data from Fig.~\ref{fig:xtip}b).

We notice that the experimental data tends to gather  under the one-to-one $\tau_{th}$-line. We attribute this divergence to the fact that in the experimental setup we do not have a sharp wedge. However,  the numerically estimated exponents for which this finite curvature radius $h_0$  is taken into account, agree with  the experimental measurements from Fig.~\ref{fig:xtip}b) within the error bar in these data, except for the
smallest and largest $\tau$-values.

At the largest
$\tau$-value (the Newtonian case) the discrepancy is still moderate and may be linked to the 
difference in the Newtonian and non-Newtonian behavior near the contact line. On the other hand, the discrepancy at the smallest $\tau$-value is likely caused by a departure from the non-Newtonian domain of the power-law fluid, which happens at sufficiently small shear rates.
In the Supplemental Material, we show how the numerical $\tau$-values depend on  $h_0$ 
and converges to the analytical result as  $h_0 \rightarrow 0$. 
Variations of the initial half-width $w_0$ in the $0-2$ mm range has less than a 1 \% effect on the $\tau$-values as initial Gaussian
profiles converge quickly to the profiles of \eq{kuyy} (see Supplemental Material).

Fig.~\ref{fig:datacollapse}a) shows the experimental profiles $h(x,t)$  for the spreading of a drop at a XG-concentration of $2\,\mathrm{g/L}$ and $V_0 = 19\,\mathrm{mm}^3$, while Fig.~\ref{fig:datacollapse}b)
shows the data collapse implied by \eq{kuykugg} and \eq{juytf} outside the domain where the corner curvature introduces a cut-off.
The numerical results capture this behavior, while also showing a close
agreement between the analytic and numerical results in the central region. Note that the collapse also verifies the scaling $x_{tip}\propto t^{\tau}$ while the agreement between the analytic and numerical results implies a very fast relaxation towards the analytical prediction.  

For the experimental determination of the  cross-sectional area $A(x,t)$ employed in Fig.~\ref{fig:datacollapse}b), a correction due to the curvature radius $h_0$ along the corner of the wedge is used so that the normalization $\int dx A(x,t) =V_0$ is verified. This is achieved by subtracting from the approximately triangular cross section of the ideal wedge a smaller triangular area that represents the empty region beneath the curved segment, see Fig.~\ref{fig:experimentalsetupcombined}d). This correction is further described in the Supplemental Material.

Having shown that the spontaneous spreading of a droplet of a power-law fluid in a wedge is governed by a non-linear diffusion equation, we have proceeded via an analytic solution  of this equation to obtain the rheological exponents of several power-law  fluids by simple experimental measurements. In classical (Fickian) diffusion, the position of the diffusing front grows in time as $x \propto t^{1/2}$. In our system, by contrast, the exponent differs from $1/2$, a hallmark of anomalous diffusion. Specifically, the spreading follows $x_{tip}\propto t^\tau$, where $\tau$ was typically found to be smaller than $0.3$ for the shear-thinning fluids tested (sub-diffusive behavior).
The  connection between the anomalous diffusion exponent $\tau$  and the fluid  rheology exponent $n$ is expressed in \eq{tau}, a key result of our work and the theoretical basis for our experiments. 

In our corner geometry the $\tau$-exponent depends sufficiently strongly on the rheological exponent $n$ so as to make it experimentally possible to determine $n$ from $\tau$. The  measured
$\tau $-values are in the range $\sim 0.16-0.42$.
By contrast, comparable experiments \cite{rafai05} that study
 xanthan gum fluids spreading on a flat surface obtain $\tau $ values that are limited to 
the range $\sim 0.05-0.1$. These values are too small, or too limited in range to allow for the determination of $n$. 
From a physical perspective this difference in sensitivity is likely linked to the difference
in the nature of the driving forces between the two geometries: In a wedge of small opening
angle the capillary pressure, which drives the flow, 
is negative and increases as $1/h$ as the
film thins down towards the tip. On a flat surface ($\alpha =\pi$), on the other hand, the driving mechanism is the moving contact line where the capillary pressure is set by some dynamic contact angle without a similar $1/h$-increase. In a sharp corner the contact line motion is primarily normal to the flow direction, at least when $|\partial h/\partial x| \ll 1$, while it is along the flow direction on a flat plane. 
In our theoretical description we have taken the capillary pressure to be governed
by the curvature in the transverse direction, ignoring the smaller contribution due to the curvature
in the flow direction. 
At the critical value $ \alpha = \pi - 2\Theta$, the fluid surface has no curvature in the  direction
transverse to the flow. As $\alpha $ is increased past this critical value 
the capillary pressure due to this curvature changes sign. This signals the transition to a different flow regime, where the capillary pressure of the sharp corner geometry is replaced by a weaker pressure. In the case of spreading on a flat substrate, the  pressure is entirely
controlled by the curvature in the flow direction.

Carrying out the experiments
for a range of different liquids, we have observed robust behavior in the sense that experimental
artifacts like inertial effects, gravity, corner curvature, pinning effects, variability in the
contact angle, and finally the crossover to Newtonian behavior at small shear rates, do not dominate in the end the measurements of the rheological exponent.

We thank Alex Hansen, Per Arne Rikvold and Erika Eiser for helpful discussions during the execution of this project. This work was partly supported by the Research Council of Norway through its projects 262644 (Center of Excellence funding scheme), 324555 (Researcher Project for Young Talent) and 301138 (NANO2021 program).

\clearpage
\twocolumn[
    \begin{@twocolumnfalse}
        \section*{\centering Supplementary Material for the paper ``Droplet spreading in a wedge: A route to fluid rheology for power-law liquids'' by Moura et al.}
        \vspace{1cm} 
    \end{@twocolumnfalse}
]

\appendix

\section{Experimental Methods}

\subsection{Preparation of the wedge, polymeric coating and plasma treatment}

The key challenge in the preparation of the experiment is ensuring that any types of perturbations to the droplet spreading are minimized. The typical main source of perturbation is contact-line pinning, which may arise from either chemical or mechanical heterogeneities in the solid surface of the plates. To reduce contact-line pinning, we applied a hydrophilic polymeric coating to the inner surfaces of the plates, followed by low-pressure plasma treatment. The coating was created using the two-part PDMS (Polydimethylsiloxane) Dow SYLGARD 184, a silicone elastomer widely used in microfluidics. The PDMS consists of a base and a curing agent, but because it is naturally hydrophobic, we added a third component, a methyl-terminated poly(dimethylsiloxane-b-ethylene oxide) from Polysciences Inc., to act as a hydrophilic agent. We targeted weight ratios of $10\!:\!1\!:\!0.2$ for the base, curing agent, and hydrophilic agent, respectively, with the hydrophilic agent incorporated while the silicone was still fluid, before curing. The mixture was stirred vigorously for several minutes, then placed in a vacuum chamber to eliminate air bubbles introduced during stirring.

While the coating solution was degassing, we prepared the glass plates by positioning them side by side along their longest edge ($76\,\mathrm{mm}$). We connected the plates with a strip of Scotch tape acting as a hinge along the edge (see Fig.~\ref{fig:experimentalsetupcombined}d). Care was taken to avoid trapping air bubbles, especially along the central edge where the droplet spreading occurs.

Next, we used a Laurell WS-650-23B spin coater to apply the coating solution to the glass slides. We added the degassed PDMS solution to the central part of the plates and spun them at 3000 RPM for 2 minutes, following a brief fast acceleration phase. The spinning was repeated once again to ensure a homogeneous thin layer of the modified PDMS solution. The sample was weighed before and after coating, revealing that approximately $0.09\,\mathrm{g}$ of PDMS had been deposited, corresponding to a layer thickness of about $0.05\,\mathrm{mm}$. For reference, the thickness of the Scotch tape beneath the PDMS was estimated at $0.03\,\mathrm{mm}$, and the glass plates themselves were $1.0\,\mathrm{mm}$ thick.

After coating, the sample was cured in an oven at $100^\circ\mathrm{C}$ for 1 hour. Once the PDMS was fully cured, we proceeded with the nitrogen plasma treatment. We used a low-pressure plasma system from Diener Electronic for this step. The sample was placed in a sealed plasma chamber, and a vacuum pump reduced the air pressure to $0.15\,\mathrm{mbar}$. A steady nitrogen flow was established to create a nitrogen-rich environment at $0.3\,\mathrm{mbar}$. Plasma was then activated and maintained at $60\%$ power for 3 minutes.

This plasma treatment was crucial for achieving hydrophilicity, as the hydrophilic agent alone was insufficient. PDMS is inherently hydrophobic, with a water-air contact angle around $\Theta = 120^\circ$ \cite{tan2010}. Before plasma treatment, we measured a contact angle of $\Theta = 75^\circ$, using a droplet of xanthan gum solution ($2\,\mathrm{g/L}$ in water). After plasma treatment, the contact angle decreased to approximately $\Theta = 20^\circ$.

\subsection{Preparation of the fluids}
The power-law fluids were prepared by mixing different concentrations of xanthan gum in water using a magnetic stirrer. As noted by Whitcomb \cite{whitcomb1977}, adding just $1\%$ xanthan gum to water can enhance its viscosity up to $100\: 000$ times at low shear rates, but this increase was limited to only 10 times at high shear rates. We started the procedure by adding $250\,\mathrm{g}$ of deionized water in a container that was placed in a magnetic stirrer at a rate high enough to form a visible vortex. The XG was gradually poured onto the side walls of the generated vortex in order to avoid the formation of large clumps.  After $1\,\mathrm{h}$, we added $0.25\,\mathrm{g}$ of Nigrosin, a strong dark blue water-soluble dye.  We also prepared a reference Newtonian fluid by simply mixing  glycerol and water, $50\%$ by weight of each part.

After mixing, microscopic air bubbles were inevitably trapped in the fluids, potentially affecting their rheology. To address this, the fluids were placed in a vacuum chamber for degassing.

\subsection{Rheological measurements using a commercial rheometer}

\begin{figure}[]
\begin{center}
\includegraphics[width=1.0\columnwidth,angle=0]{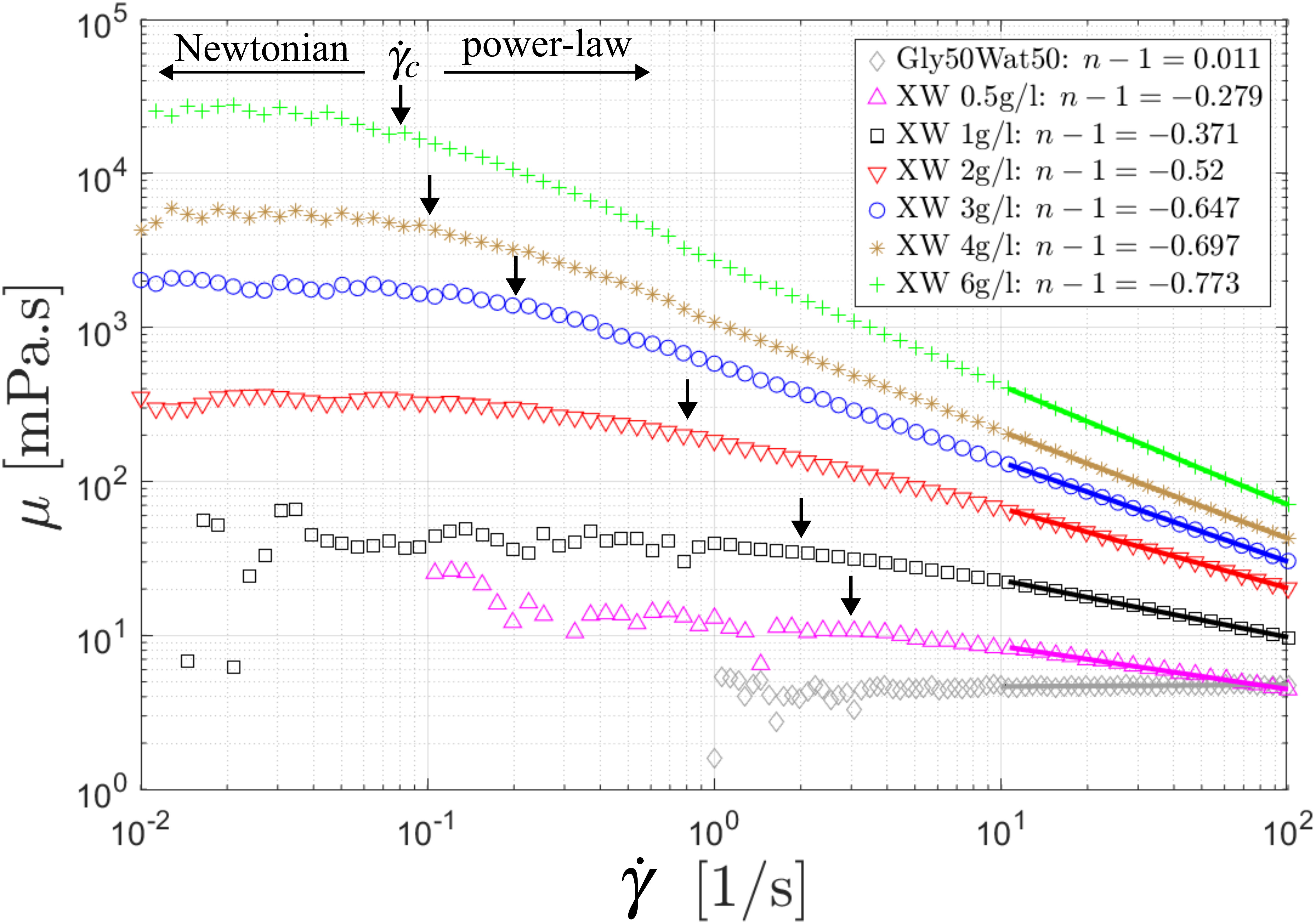}
\caption[]{Viscosity curve from the Anton Paar MCR 702e rheometer for the different power-law fluids. The reference Newtonian glycerol-water mixture  is also shown. The region used to extract the power-law exponent $n-1$ corresponds to the solid lines. The  shear rate $\dot{\gamma}_c$, above which the power-lar rheology is observed, is marked by black arrows.}
\label{fig_rheology}
\end{center}
\end{figure}

The rheology of all fluids was characterized using an Anton Paar rheometer model MCR 702e. The results for the effective viscosity $\mu$ as a function of the shear rate $\dot{\gamma}$ are shown in Fig.~\ref{fig_rheology}. We conducted tests in which the shear rate $\dot{\gamma}$ was gradually increased from $\dot{\gamma}_{\mathrm{min}} = 0.01\,\mathrm{s}^{-1}$ to $\dot{\gamma}_{\mathrm{max}} = 100\,\mathrm{s}^{-1}$ for XG fluids with concentrations between $1$ and $6\,\mathrm{g/L}$; from $\dot{\gamma}_{\mathrm{min}} = 0.1\,\mathrm{s}^{-1}$ to $\dot{\gamma}_{\mathrm{max}} = 100\,\mathrm{s}^{-1}$ for the XG fluid at $0.5\,\mathrm{g/L}$; and from $\dot{\gamma}_{\mathrm{min}} = 1\,\mathrm{s}^{-1}$ to $\dot{\gamma}_{\mathrm{max}} = 100\,\mathrm{s}^{-1}$ for the Newtonian glycerol--water mixture. The higher minimum shear rates used for the two least viscous fluids were necessary because the torque at low shear rates was too small to be reliably measured by the rheometer. Note that all fluids present a power-law regime behavior of the form $\mu \propto \dot{\gamma}^{(n-1)}$ for high values of the shear rate $\dot{\gamma}$ (shown as solid lines in the figure). The fluids became more shear-thinning for higher concentrations of XG, see the concentration and exponent values in the legend. The power-law regime is only observed beyond some threshold value $\dot{\gamma}_c$, which depends on the concentration of XG and is indicated by the black arrows. Increasing the XG concentration causes $\dot{\gamma}_c$ to decrease. When $\dot{\gamma}<\dot{\gamma}_c$, the fluids are in a crossover region and for very small shear rates the  fluids present a Newtonian behavior with a constant viscosity. Measurements at extremely high shear rates were not performed, however, at such rates, the viscosity of xanthan gum solutions tends to approach that of water \cite{phillips2009}. This crossover regime is well documented \cite{schowalter1978} and was captured in more complex models for the viscosity curve, such as that by Carreau \cite{carreau1972}. 

The $\mu_0$ values that may be read from Fig.~\ref{fig_rheology} at $\dot{\gamma}_0 = 10\,\mathrm{s}^{-1}$ are roughly within a factor 2 of the values obtained from \eq{eq:eta0},  using the static contact angle. However, if this value is replaced by a dynamic contact angle $\Theta_{dyn}$, agreement between the values obtained from Fig.~\ref{fig_rheology} and the prediction  of  \eq{eq:eta0} may be acheived, using 
$\Theta_{dyn} \approx \pi / 3$ as a fitting parameter.
This value agrees well with the $\Theta_{dyn} \approx 65^\circ$ obtained by Wijnhorst et al. \cite{wijnhorst20} for Newtonian fluids spreading in an $\alpha =\pi/2$ wedge. 
This strongly suggests that, while 
the theoretical exponent $n$ obtained from the measured $\tau$-values is robust in the sense
that it does not depend much on the idealizations made in the theory, the prefactor $\mu_0$ does.

In Fig.~\ref{fig_rheology}, we also notice that for the samples presenting the lower values of the effective viscosity, the data begins to fluctuate and becomes unreliable for very low shear rates. We believe this happens because in this zone, the torque imposed by the fluid on the rheometer plates is too small to be reliably measured. This is clearly observed in the curves for XG at $1\,\mathrm{g/L}$, where fluctuations appear below $\dot{\gamma} = 0.04\,\mathrm{s}^{-1}$; for XG at $0.5\,\mathrm{g/L}$, where fluctuations begin below $\dot{\gamma} = 0.3\,\mathrm{s}^{-1}$; and for the Newtonian case, where they start around $\dot{\gamma} = 3\,\mathrm{s}^{-1}$. In this study, we restrict our analysis to the portion of each curve that lies within the power-law regime.

\subsection{Droplet placement and effects from films on the plates}

A micrometer syringe with a flat-tipped needle is used to deposit a droplet of the specified liquid at the wedge's center. The external diameter of the needle is $1.80\,\mathrm{mm}$ and the droplet volume (set on the syringe) is approximately $20\,\mu\mathrm{L}$. The micrometric needle is positioned in an arm connected to a translation stage which allowed for $x$, $y$ and $z$ translation to position the droplet in the wedge. The procedure of placing one droplet consisted in first generating the pendant droplet on the needle and then slowly moving the translation stage down until the droplet touched the inner sides of the plates. It would then disconnect from the needle and start spreading sideways, see Fig.~\ref{fig:experimentalsetupcombined}. Notice that once the droplet touches the glass plates, it creates a wetted region close to the center, which persists as the experiment progresses. This can be seen on Fig.~\ref{fig:experimentalsetupcombined}c) as the darker patch just below the tip of the needle in the central part of the image. The presence of this film can cause the droplet shape for small $x$ values (close to the center) to deviate from the theoretically predicted shapes, where this artifact was not present. The lateral extent of this region is of the order of $0.3\,\mathrm{cm}$ in both directions, so in our analysis we have ignored the initial frames of the dynamics, in which the droplet spreading may be more affected by these films.

\subsection{Experiments with highly wetting silicone oils using a simpler wedge construction}
\begin{figure}[t]
\begin{center}
\includegraphics[width=\columnwidth,angle=0]{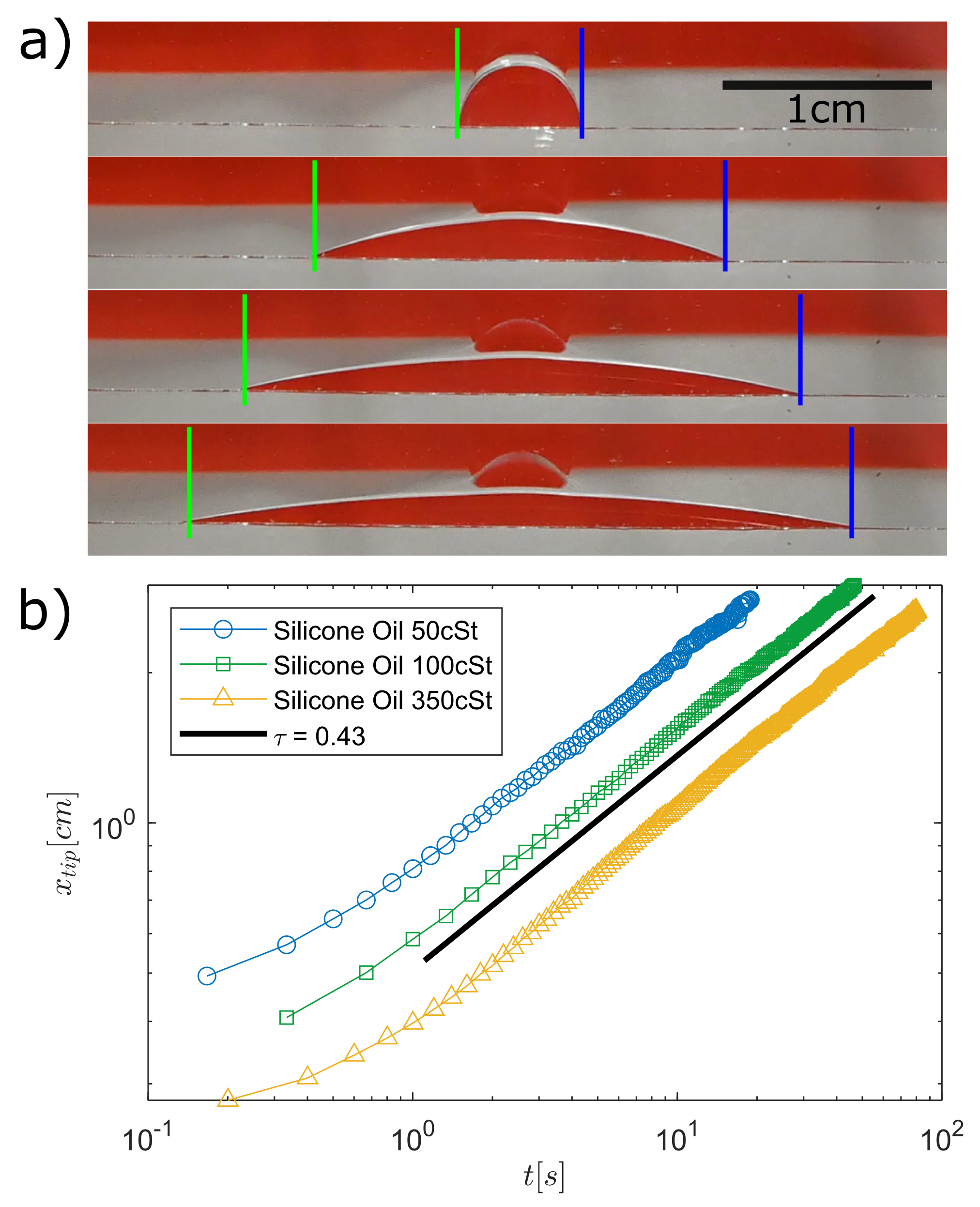}
\caption[]{a) Snapshots showing the evolution of a droplet of highly wetting silicone oil with a viscosity 100 cSt in the  wedge composed only of  two glass slides. The interval between the top and bottom snapshots is $\Delta t = 10\,\mathrm{s}$. b) The spreading dynamics of three silicone oils with viscosities $50\,\mathrm{cSt}$, $100\,\mathrm{cSt}$, and $350\,\mathrm{cSt}$ show similar exponents $\tau = 0.43 \pm 0.03$,
(black line), consistent with the theoretical value $\tau = 0.4$ predicted for a Newtonian fluid. The data is shifted horizontally by different values to aid visualization.}
\label{fig:silicone}
\end{center}
\end{figure}

For highly wetting liquids, such as silicone oils, contact-line pinning is naturally reduced, and a simpler setup with only the glass wedge without any polymeric coating or plasma treatment is sufficient. As a first step in the development of our technique, we performed experiments with this simpler wedge construction using silicone oils with kinematic viscosities $50\,\mathrm{cSt}$, $100\,\mathrm{cSt}$, and $350\,\mathrm{cSt}$. In Fig.~\ref{fig:silicone}a) we show snapshots of the experiment with the 100 cSt silicone oil. The transparent silicon oil appears red because of a convenient lensing effect: A red tape is positioned behind the wedge and the contrast in refracting indices causes the droplet to refract the red light into the camera, thus avoiding the need for dyes. The interval between the first and last snapshots was $\Delta t = 10\,\mathrm{s}$.  The spreading dynamics is seen in Fig.~\ref{fig:silicone}b) where we see that all exponents show values  $\tau = 0.43 \pm 0.03$, obtained from the black guide-to-the-eye line. These findings are consistent with the fact that silicone oil is a Newtonian fluid with a theoretical value  $\tau = 0.4$, which results from setting $n=1$ in  \eq{tau}. In this experiment, the wedge had an opening angle of $\alpha = 20^\circ$, and the droplet was positioned by hand instead of employing the more precise arrangement with a micrometric syringe. Images were recorded with a DSLR camera in video mode.  Notice that, even though the fluids are significantly different, with viscosities varying by a factor of 7, the exponent is rather stable. The specific value of the viscosity, as well as the opening angle $\alpha$, enters in the prefactor governing the scaling through the `diffusion coefficient' $D_0$ in \eq{eq:D0} but do not affect the exponent $\tau$.

\subsection{Estimation of cross-sectional area of droplet from experimental height profile}

In the experiments the cross-sectional area $A(x,t)$ is affected by  the curvature of the edge as shown schematically in Fig.~\ref{fig:experimentalsetupcombined}d). If we ignore this effect, the computed volume of the droplet differs from the known value $V_0$, which is injected by the  needle. The effective height $h$ measured experimentally is the vertical distance from the top of the liquid-air interface and the curved segment in Fig.~\ref{fig:experimentalsetupcombined}d), which itself is at a distance $\approx h_0$ from the extrapolated, sharp edge of the wedge. We can then estimate $A(x,t)$ as the difference between the area of the triangular wedge (ideal case with sharp corner) and the void space under the curved segment seen in Fig.~\ref{fig:experimentalsetupcombined}d). From a simple geometrical argument we get

\begin{equation}
    A \approx (h+h_0)^2\tan\left(\frac{\alpha}{2}\right)-h_0^2\tan\left(\frac{\alpha}{2}\right) = (h^2 +2hh_0)\tan\left(\frac{\alpha}{2}\right) \:.
\end{equation}

The vertically projected effective height $h$ is what is shown in Fig.~\ref{fig:datacollapse}a). Notice that this is slightly different from $h$ seen in Fig.~\ref{fig:experimentalsetupcombined} but can be obtained from the latter as $h \cos(\alpha/2)-h_0$.

\section{Numerical model}  

\begin{figure}[]
\begin{center}
\includegraphics[width=0.8\columnwidth,angle=0]{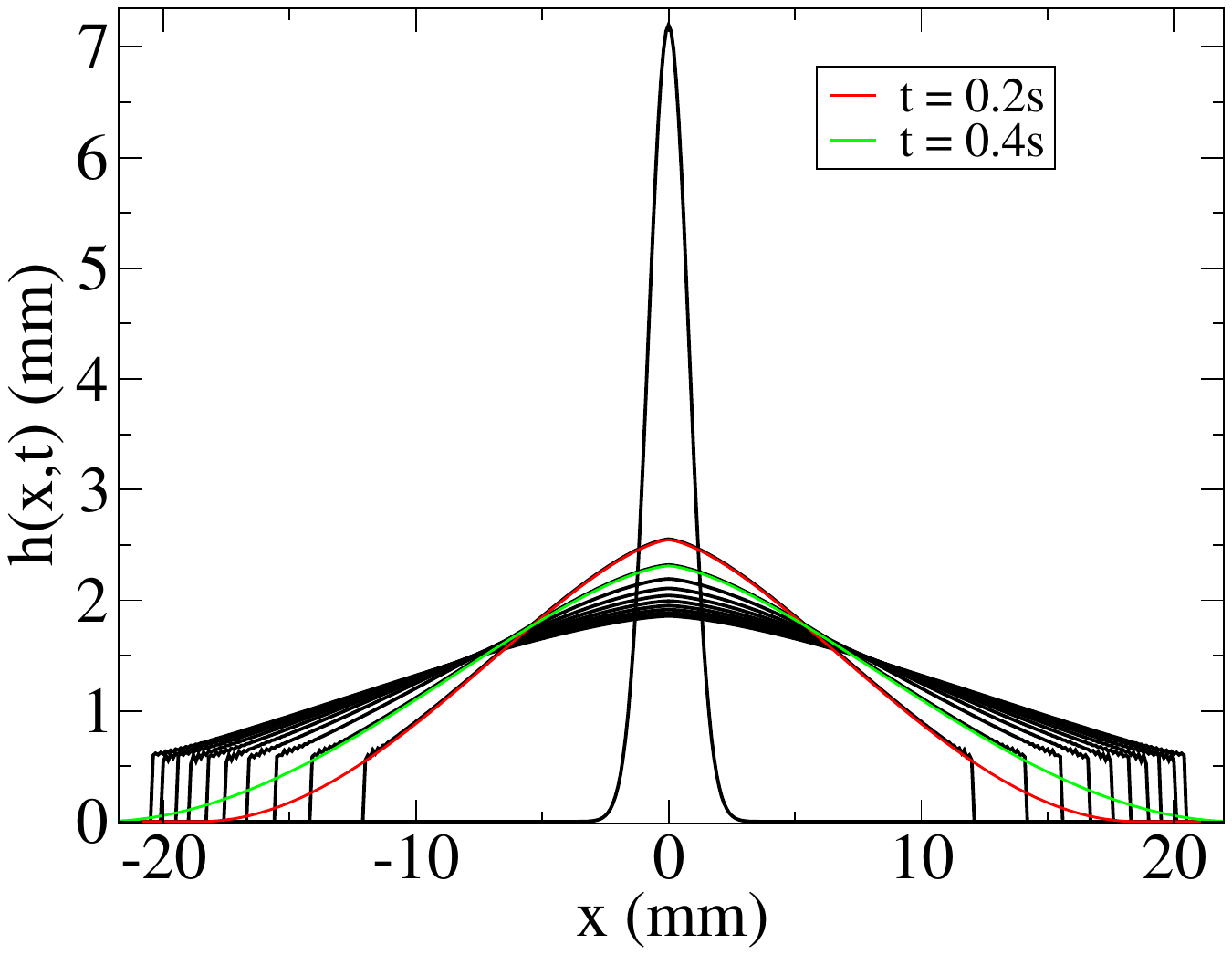}
\caption[]{The film height as a function of position along the
  corner at different times when the rheological exponent $n=0.48$.
  The black lines derive from  the numerical solutions of
  \eq{conserved2}
  using $h(x,t)=\sqrt{2A(x,t)/\alpha}$, while the colored
  lines show the corresponding analytical solution of \eq{kuykugg} and
  \eq{kuyy} for two different times.  The taller central curve shows the initial profile of the droplet. A volume of $19\,\mathrm{mm}^3$, initial half-width $w_0 = 0.8\,\mathrm{mm}$, and corner radius of $h_0 = 0.6\,\mathrm{mm}$ were used as parameters for all curves.}
\label{fig2}
\end{center}
\end{figure}
\begin{figure}[t]
\begin{center}
\includegraphics[width=0.66\columnwidth,angle=0]{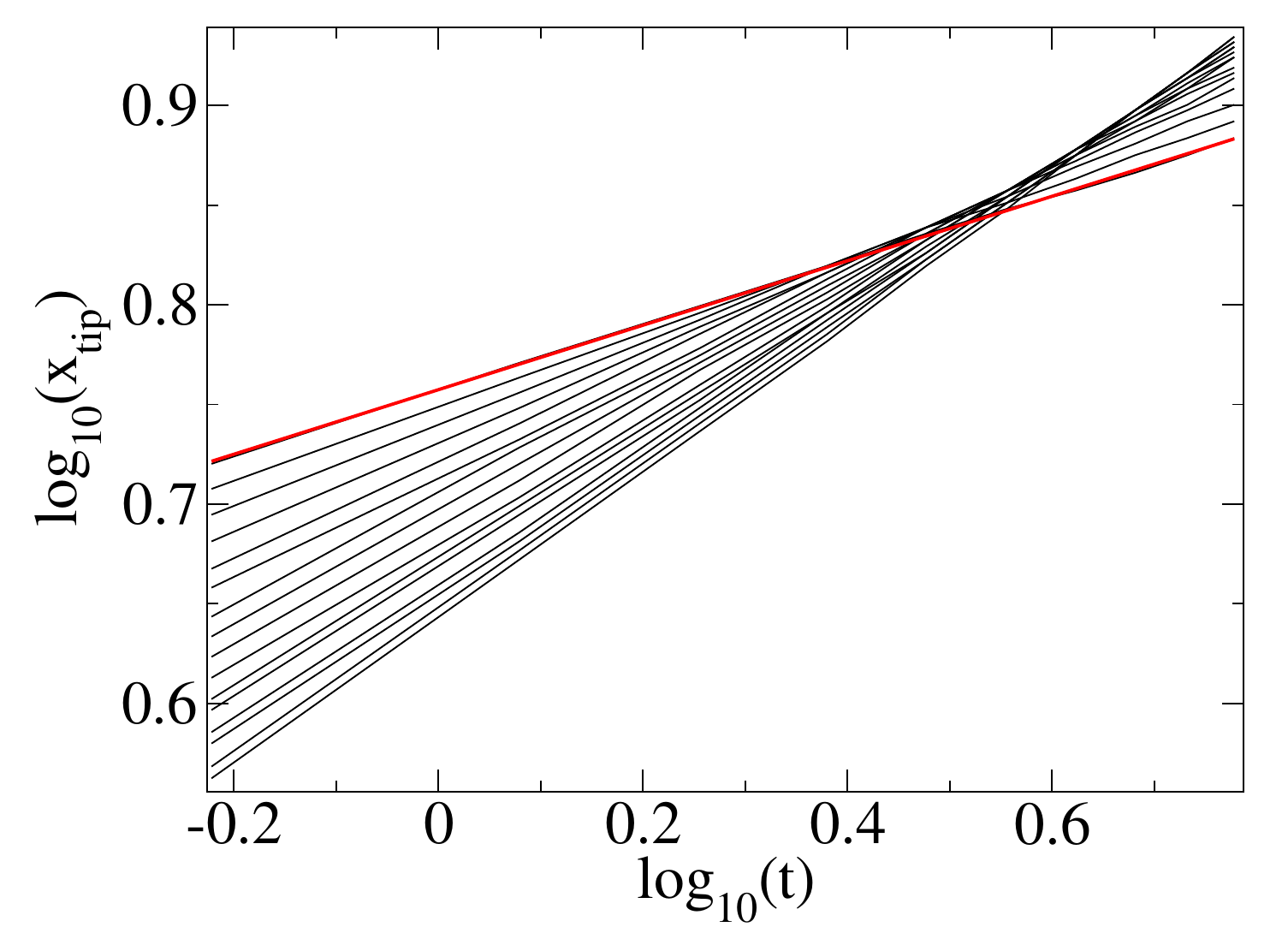}
\caption[]{The position of the right droplet tip  as a function of
  time for $n$-values in the range $0.25-1.0$ (black curves). The red
  curve is a linear fit to the black $n=0.25$ curve, which it nearly covers.  All other parameters are
  as in Fig.~\ref{fig2}. }
\label{fig3}
\end{center}
\end{figure}
\begin{figure}[t]
\begin{center}
\includegraphics[width=0.66\columnwidth,angle=0]{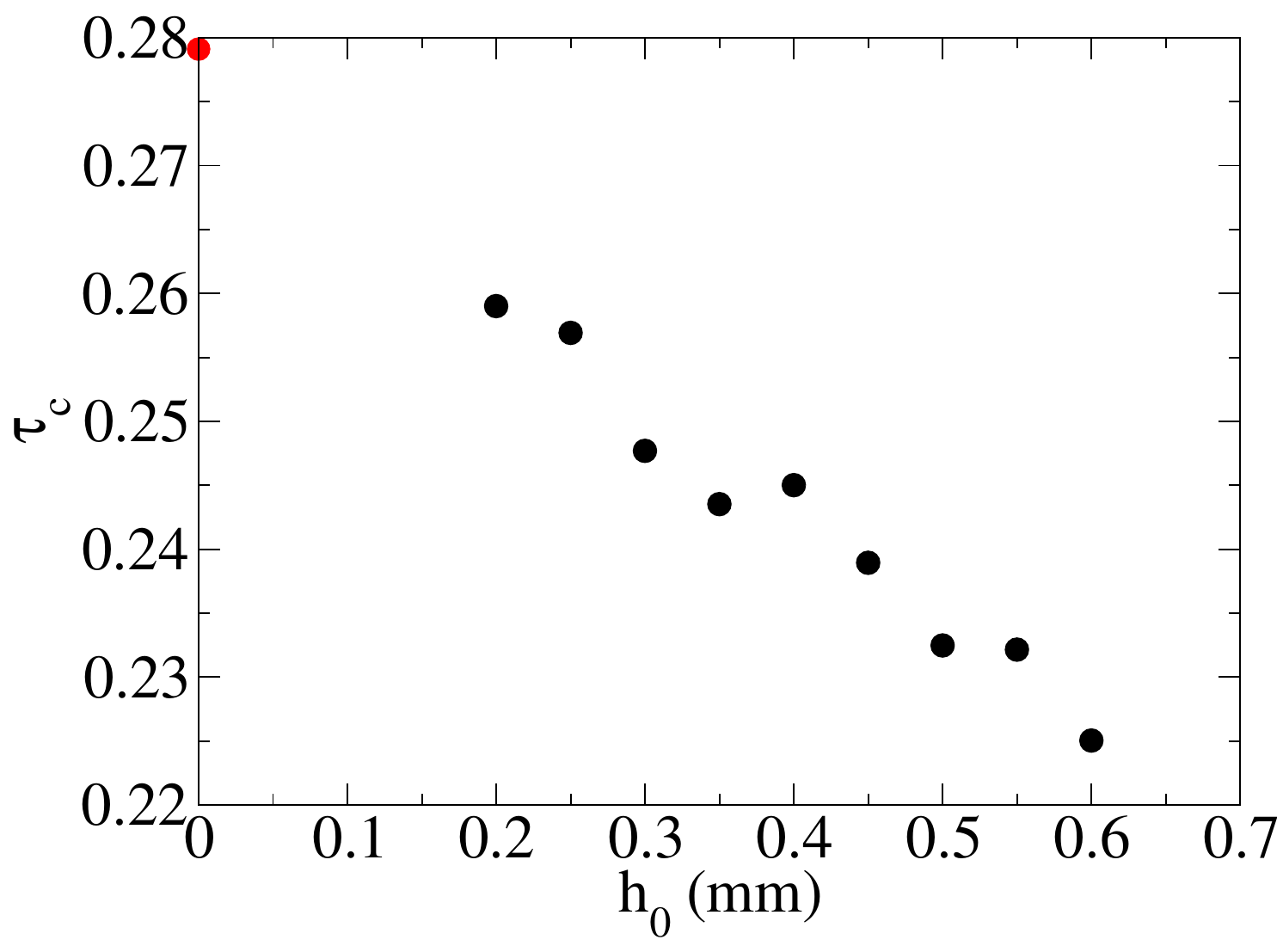}
\caption[]{The measured slopes $\tau_c$ derived from the numerical 
  solution of \eq{conserved2} versus the radius of curvature of the corner $h_0$. Here $n=0.48$ and the red dot shows the theoretical value  given in   \eq{tau}. }
\label{fig5}
\end{center}
\end{figure}

 In Fig.~\ref{fig2} the droplet height $h(x,t) = \sqrt{2A(x,t)/\alpha}$
is plotted as a function of
position,  where we have taken the normalization to
be $\int dx\, A(x,t) = 19\,\mathrm{mm}^3$, approximating the experimental value.
Note that the analytic solution (red and green curves for 
times given in the legend) has a cusp at $x=0$ where the second
spatial derivative of $h(x,t)$ diverges, while the numerical solution
is initialized with the smooth  Gaussian $\propto \exp(-x^2/w_0^2)$ (shown as the taller curve).
Nevertheless, the numerical solution converges to the analytical solution (red curve)  with the same droplet volume within 1‰ during the first 1/10 of the time span, except at the cut-off point $h\approx
h_0$, where a weak numerical oscillation around the analytic solutions causes deviations of the order 1\%. This shows that the 
 numerical solution quickly converges to the analytical shape prediction, except for the values where $h$ is below the $h_0$-cutoff.
Figure~\ref{fig3} shows that $x_{tip}(t)$, when  $h_0 = 0.6\,\mathrm{mm}$ and $w = 0.8\,\mathrm{mm}$, is well approximated by a power law,
 that is, that it yields straight lines on a log-log plot. 
In Fig.~\ref{fig5} the slopes of these plots are given, showing 
the variation of $\tau_c$ with the corner curvature $h_0$. Note that they extrapolate well to the analytic value
given by the red dot. The deviation from the theoretical $\tau$ value decreases with decreasing $h_0$, as expected.

\section{Replacement of Darcy's law for a power-law fluid}  
We need the relationship between the pressure gradient and the average 
flow velocity of a power-law fluid, where the viscous stress scales as $ \dot{\gamma}^n$, where $\dot{\gamma}$ is the shear rate, and
$n$ is the rheological exponent, which for a shear-thinning fluid is smaller than unity. When $n=1$ the fluid
is Newtonian.
Just as the viscous stress in a Newtonian fluid, the
non-Newtonian fluid stress tensor can only depend on the symmetric part of the
velocity gradient. For an incompressible fluid the total stress tensor, which includes the
pressure $P$,  may be cast in the
covariant form,
\be
\sigma_{tot\;ij}
= - \delta_{ij} P + \eta_0 \dot{\gamma}_{ij} (\dot{\gamma}_{kl}\dot{\gamma}_{kl})^{(n-1)/2} \:,
\ee
where $P$ is the pressure, $\delta_{ij}$ the Kronecker delta function, and summation over
repeated indices is implied. 
 The viscosity coefficient (in units of $\mathrm{Pa\,s}^n$) may be decomposed 
 into $\eta_0 = \mu_0 \dot{\gamma}_0^{1-n} $, where $\mu_0  $ (in units of $\mathrm{Pa\,s}$) is the  viscosity at the shear rate $\dot{\gamma}_0$. 
The  strain rate tensor is given by
\be
\dot{\gamma}_{ij} =  \frac{\partial u_i}{\partial x_j}  + \frac{\partial u_j}{\partial x_i} \:,
\ee
and the stress balance equation reads
\be
\nabla \cdot \sigma_{tot} =0\:.
\label{kug}
\ee
When  $n=1$, this  equation  reduces to the Stokes equation.

Introducing a characteristic length $h$, which could be the channel
width,  this equation may be written in terms of the non-dimensional
primed quantities defined through
\begin{align}
x_i &= h x'_i\nonumber  \\
u_i &= \dot{\gamma}_0 h u'_i\nonumber \\
      \dot{\gamma}_{ij}
  &=    \dot{\gamma}_0  \dot{\gamma}'_{ij} \nonumber \\
    \nabla
  &= \frac{1}{h}   \nabla' \:.
\end{align}
Writing the pressure gradient as
\be 
\nabla P = -|\partial P/\partial x| {\bf e}\:,
\ee 
where ${\bf e}$ is the unit vector in the opposite direction of the pressure 
gradient, \eq{kug} may be written as
\be
{\bf e}= -\nabla' \cdot \left(  \dot{\tilde{\gamma}} \left(
\dot{\tilde{\gamma}}^2
  \right)^{(n-1)/2} \right) \:,
\label{kuyf}
\ee
where $ \tilde{\gamma} =  \gamma'/G^{1/n}$ and we have introduced the
dimensionless
ratio between pressure forces and viscous forces, and
\be
G=\frac{|\partial P/\partial x|h}{\mu_0  \dot{\gamma}_0} 
\ee
which is thus  a non-dimensionalized pressure gradient.
Since the boundary conditions may be given in terms of $\bu' (\bx'_B
)$ where $\bx_B$ are boundary coordinates, the
stress balance equation yields a solution
of the form
\be 
\tilde{\gamma}_{ij}= f_{ij} (\bx') \:,
\ee
which in turn gives
\be 
\gamma'_{ij}= f_{ij} (\bx')  G^{1/n} \sim
\frac{\partial
  u'_i}{\partial x'_j}\:.
\ee
Integrating this equation for the velocity component along the wedge gives
\be 
u' =  G^{1/n} F_n (\bx' )\:, 
\ee
for some other dimensionless function $F_n$,
which in turn gives
\be 
u =  G^{1/n} \dot{\gamma}_0 h  F_n (\bx' ) \:. 
\ee
Averaging over a cross section of the flow gives
\be
\overline{u} =
G^{1/n} \dot{\gamma}_0 h Q_n
= \left( \frac{|\partial P/\partial x| }{{\eta_0}} \right)^{1/n}  h^{1+1/n}Q_n\:.
\ee

It is possible to calculate $Q_n$ in the lubrication approximation, assuming that the flow is governed by the gradients of $u$ in the angular direction (normal to the wedge walls) alone.

We shall start with the velocity field $u(x)$ in the $z$-direction in  a straight channel of half-width $a$ and
 a coordinate $x$ in the direction transverse to the flow where $x=0$ in the middle. The velocity satisfies
 the boundary condition $u(\pm a)=0$. Then,  \eq{kug} reduces
to 
\be
2^{\frac{1-n}{2}} \frac{|\partial P/\partial x|}{\eta_0}
 - \partial_x ((-\partial_x u)^{n})=0 
\ee
for $x>0$.  This equation is easily integrated to yield 
\be
u(x) = 2^{\frac{1-n}{2n}}\frac{n}{n+1} \left( \frac{|\partial P/\partial x|}{\eta_0}\right)^{1/n} ( a^{\frac{n+1}{n}} -x^{\frac{n+1}{n}} )\:,
\ee
which has the cross-sectional average, 
\be
\overline{u} = \frac{1}{a}\int_0^a dx u(x) =
2^{\frac{1-n}{2n}}\frac{n}{2n+1} \left( \frac{|\partial P/\partial x|}{\eta_0}\right)^{1/n}  a^{\frac{n+1}{n}} \:.
\ee
This expression agrees with Darcy's law in a channel of
half-width $h$ in the Newtonian  $n=1$ case where
$\overline{u} =  h^2 |\partial P/\partial x|/(3 \eta_0)$.

Now,  we return to the wedge geometry. Neglecting the effects of transverse flow on the pressure, the pressure gradient points in the $x$-direction and is constant over any fixed $x$ plane inside the  fluid. 
For a narrow wedge we may therefore approximate the flow as a superposition of channel flows with half-widths $\alpha r/2$.  This approximation is limited to small $\alpha$- values as it assumes that the velocity variations are dominantly in the angular direction and negligible in the radial direction.  So, we replace $a\rightarrow \alpha r/2$, thereby obtaining a lubrication 
approximation $\overline{u}(r)$ to the  flow at a distance $r$ to the corner. Averaging $\overline{u}(r)$ over the cross-sectional area 
\be
\langle u \rangle = \frac{2}{ h^2} \int_0^h drr \overline{u}(r)
\ee
yields the volume flux
\be
q = \frac{\alpha h^2}{2}\langle u \rangle = Q_n \left( \frac{|\partial P/\partial x|}{\eta_0}\right)^{1/n}h^{3+1/n} \:,
\ee
where
\be
Q_n = \frac{2^{\frac{1+n}{2n}} n^2}{(2n+1)(3n+1)} \left( \frac{\alpha}{2} \right)^{\frac{2n+1}{n}},
\ee
which is the desired result.

\end{document}